\documentclass[11pt]{article}
\usepackage{times}
\usepackage{a4wide}
\usepackage{amsfonts}
\usepackage{amssymb}
\usepackage{multicol}
\usepackage{amsmath}
\usepackage{graphicx}
\usepackage[usenames,dvipsnames]{xcolor}
\usepackage{ifpdf}
\ifpdf
\usepackage[pdftex,unicode,implicit]{hyperref}

\DeclareSymbolFont{extraup}{U}{zavm}{m}{n}
\DeclareMathSymbol{\varheart}{\mathalpha}{extraup}{86}
\DeclareMathSymbol{\vardiamond}{\mathalpha}{extraup}{87}

\usepackage[sort&compress,comma]{natbib}
\bibpunct{[}{]}{,}{n}{}{,}

\hypersetup{%
  pdftitle    = { },
  pdfkeywords = {string theory, black holes, Calabi Yau, compactification, quantum corrections, supergravity model, Lambert function},
  pdfauthor   = {Pablo Bueno and C. S. Shahbazi},
  plainpages  = true,
  colorlinks  = true,
  citecolor   = Red	,
  urlcolor    = Green,
  linkcolor   = Blue	
}

\else
  \usepackage[dvips]{graphicx}
  \usepackage[unicode,implicit]{hyperref}

\fi

\makeatletter
\renewcommand\section{\@startsection{section}{1}{\z@}%
                                  {-3ex \@plus -1ex \@minus -.2ex}%
                                  {2ex \@plus.2ex}%
                                  {\small\small\bfseries}}
\makeatother

\makeatletter
\renewcommand\subsection{\@startsection{subsection}{1}{\z@}%
                                  {-3ex \@plus -1ex \@minus -.2ex}%
                                  {2ex \@plus.2ex}%
                                  {\small\small\bfseries}}
\makeatother

\makeatletter
\@addtoreset{equation}{section}
\makeatother

\begin{document}

  \begin{@twocolumnfalse}

\begin{center}
{\large{\bf \textcolor{Black}{Non-perturbative black holes in Type-IIA String Theory vs. the No-Hair conjecture}}}

\begin{center}

\renewcommand{\thefootnote}{\alph{footnote}}
{\normalsize \textcolor{Black}{Pablo~Bueno}$^{\textcolor{Red}{\varheart}}$
}
{\normalsize and \textcolor{Black}{C.~S.~Shahbazi}$^{\textcolor{Purple}{\varheart}}$
}

\renewcommand{\thefootnote}{\arabic{footnote}}
\vspace{0.1cm}

 \small{ IFT UAM/CSIC, Madrid }\\

\end{center}

\small{{\bf Abstract}}

\begin{quotation}

We obtain the first black hole solution to Type-IIA String Theory compactified on an arbitrary self-mirror Calabi Yau manifold in the presence of non-perturbative quantum corrections. Remarkably enough, the solution involves multivalued functions, which could lead to a violation of the No-Hair conjecture. We discuss how String Theory forbids such secenario. However the possibility still remains open in the context of four-dimensional ungauged Supergravity.

\end{quotation}
\end{center}

  \end{@twocolumnfalse}

\setcounter{footnote}{0}

\pagestyle{plain}



\small{\section*{Introduction}}
\noindent
Black hole physics is an extremely active research field in String Theory. Some impressive results have been obtained towards a complete match of the microscopic and the macroscopic entropies of extremal black holes \cite{Strominger:1996sh} beyond leading order, see \cite{Sen:2007qy,Mandal:2010cj} and references therein. So far, most of the literature has been focused on the so called \emph{higher-order} (curvature) corrections \cite{LopesCardoso:1998wt,LopesCardoso:1999ur,LopesCardoso:2000qm,Mohaupt:2000mj,Dabholkar:2004dq}. These modify the usual Bekenstein-Hawking area law

\begin{equation}\notag
S=\frac{A}{4}\, ,
\end{equation}

\noindent
and are prescribed by String Theory to appear in the effective classical Supergravity description. In the context of Type-IIA, the effective theory is described by a two-derivative Supergravity at tree level in $g_s$, with the higher-order corrections ocurring already at the 1-loop level.

There exist, however, a different type of \emph{stringy} corrections (which correct the point particle behaviour), some of which do not modify the effective Lagrangian of Type-IIA String Theory with higher order terms, but modify the couplings and the scalar manifold geometry of the classical Supergravity action. These are the $\alpha^{\prime}$-corrections\footnote{Perturbative (world-sheet loops) and non-perturbative (world-sheet instantons).}, often referred to as \emph{quantum} corrections, at least in the context of Type-IIA Calabi-Yau (C.Y.) compactifications \cite{Candelas:1990qd,Candelas:1990rm,Hosono:1993qy,Hosono:1994av}. In contradistinction to the curvature ones, these do not modify the Bekenstein-Hawking area law\footnote{They obviously modify the area of the black hole itself, as far as the structure of the solution is changed when they are taken into account.}.

It turns out that not so much attention has been paid to the effects of these \emph{quantum} corrections to black hole solutions of the classical Supergravity action. Some exceptions, in which this kind of solutions (or their corresponding critical points) in the presence of these have been considered, are \cite{Behrndt:1997ei,Gaida:1997id,Behrndt:1997gs,Gaida:1998pz,Galli:2012pt,Bellucci:2007eh,Bellucci:2009pg}. 

In \cite{Bueno:2012jc}, using the H-FGK formalism\footnote{A related formalism can be found in \cite{Mohaupt:2010fk,Mohaupt:2011aa}.} \cite{Galli:2011fq,Meessen:2011aa,Meessen:2012su,Galli:2012ji}, a new class of black holes for Type-IIA C.Y. compactifications was defined: they exist only when the perturbative corrections to the prepotential are included and no classical limit can be assigned to them. They were called, in consequence, \emph{quantum} black holes. Therefore, for self-mirror C.Y. manifolds such black holes do not exist, since in such case the perturbative corrections exactly vanish. However, the situation can be changed if we add non-perturbative corrections to the prepotential. That is the case we are going to consider in this letter. We will obtain the first explicit black hole solution of Type-IIA String Theory compactified on a self-mirror Calabi-Yau threefold in the presence of non-perturbative corrections, proving at the same time that these non-perturbative corrections \emph{lift} the singular behaviour of the quantum black holes to a regular one.

The solution (general class of solutions, in fact) may allow also for an explicit check of the match between the microscopic String Theory entropy and the macroscopic entropy of a supersymmetric black hole solution in the presence of these non-perturbative corrections.  
 To that respect, some partial results can be found in \cite{Behrndt:1997ei}. In any case, computing the macroscopic solution is undoubtedly the first step towards the resolution of this puzzle. 

Somewhat surprisingly, we obtain a class of solutions which involves Lambert's W function \cite{Lambert}, which is multi-valued in a certain real domain. We will explain how this fact seems to provide an appropriate scenario for a potential violation of the (corresponding uniqueness conjecture, and as a consequence of the) No-Hair conjecture in four dimensions. It turns out that, in our set-up, String Theory forbids the use of the Lambert function to that end. However, the possibility remains open in an exclusive Supergravity set-up (not necessarily embedded in String Theory) \cite{POS}, and there does not seem to be a reason to discard it right away. Black holes in ungauged four-dimensional Supergravity still hold some surprises \cite{Galli:2012jh,Hristov:2012nu}, which indicate that we may not completely understand them even in the extremal case.

Finally, it would also be of major interest to extend the construction of quantum black holes to the case of multicentered configurations \cite{Denef:2007vg,Cacciatori:2012tj,Ferrara:2012qm,Ferrara:2012yp,Andrianopoli:2011gy,Ferrara:2010ug,Denef:2000nb,Bates:2003vx,Goldstein:2008fq,Bena:2009en,Bossard:2011kz,Jejjala:2005yu,Bena:2009qv,Bobev:2009kn}.\\


\section{Type-IIA String Theory on a Calabi--Yau manifold}
\label{sec:generaltheory}


Type-IIA String Theory compactified to $4$D on a C.Y. three-fold, with Hodge numbers $(h^{1,1},h^{2,1})$, is described, up to two derivatives, by a $\mathcal{N}=2, d=4$ Supergravity whose prepotential is given in terms of an infinite series around $\Im{\rm m}z^i\rightarrow \infty$
\cite{Candelas:1990pi,Candelas:1990rm,Candelas:1990qd}
\begin{equation}
\label{eq:IIaprepotential}
\mathcal{F} =  -\frac{1}{3!}\kappa^{0}_{ijk} z^i z^j z^k +\frac{ic}{2}+\frac{i}{(2\pi)^3}\sum_{\{d_{i}\}} n_{\{d_{i}\}} Li_{3}\left(e^{2\pi i d_{i} z^{i}}\right) \ \, ,
\end{equation}

\noindent
where $z^i,~~ i =1,...,n_v+1=h^{1,1}$, are the scalars in the vector multiplets. There are also $h^{2,1}+1$ hypermultiplets in the theory. However, they can be consistently set to a constant value (See \cite{Shahbazi:2013ksa}, section 3.2). $c=\frac{\chi\zeta(3)}{ (2\pi)^3}$ is a model-dependent number, being $\chi$ the Euler characteristic, which for C.Y. three-folds is given by $\chi=2(h^{1,1}-h^{2,1})$. $\kappa^{0}_{ijk}$ are the classical intersection numbers, $d_{i}\in\mathbb{Z}^{+}$ is a $h^{1,1}$-dimensional summation index and $Li_{3}(x)$ is the third polylogarithmic function, defined in appendix \ref{sec:polylog}. The first two terms in the prepotential correspond to tree level and loop perturbative contributions in the $\alpha^{\prime}$-expansion, respectively

\begin{equation}
\label{eq:IIapert}
\mathcal{F}_{\text{P}} =  -\frac{1}{3!}\kappa^{0}_{ijk} z^i z^j z^k +\frac{ic}{2}\ \, ,
\end{equation}

\noindent
whereas the third term accounts for non-perturbative corrections produced by world-sheet instantons. These configurations get produced by (non-trivial) embeddings of the world-sheet into the C.Y. three-fold. The holomorphic mappings of the genus 0\footnote{Genus $\geq 1$ instantons contribute with higher-derivative corrections.} string world-sheet onto the $h^{1,1}$ two-cycles of the C.Y. three-fold are classified by the nubers $d_i$, which count the number of wrappings of the world-sheet around the $i-$th generator of the integer homology group $H_2(\text{C.Y.},\mathbb{Z})$. The number of different mappings for each set of $\{d_i\}$ $\left( \equiv \{   d_1,...,d_{h^{1,1}}\} \right)$ or, in other words, the number of genus $0$ instantons is denoted by $n_{\{d_i\}}$\footnote{See, e.g. \cite{Mohaupt:2000mj} for more details on the stringy origin of the prepotential.}

\begin{equation}
\label{eq:IIanonpert}
\mathcal{F}_{\text{NP}} = \frac{i}{(2\pi)^3}\sum_{\{d_{i}\}} n_{\{d_{i}\}} Li_{3}\left(e^{2\pi i d_{i} z^{i}}\right) \ \,.
\end{equation}

\noindent
The full prepotential can be rewritten in homogeneous coordinates $\mathcal{X}^{\Lambda}$, $\Lambda=(0,i)$ as
\begin{align}
\label{eq:IIaprepotentialX}
F(\mathcal{X}) =  -\frac{1}{3!}\kappa^{0}_{ijk} \frac{\mathcal{X}^i \mathcal{X}^j \mathcal{X}^k}{\mathcal{X}^0} +\frac{ic(\mathcal{X}^0)^2}{2}+\frac{i(\mathcal{X}^0)^2}{(2\pi)^3}\sum_{\{d_{i}\}} n_{\{d_{i}\}} Li_{3}\left(e^{2\pi i d_{i} \frac{\mathcal{X}^{i}}{\mathcal{X}^0}}\right) \ \, ,
\end{align}

\noindent
with the scalars $z^i$ given by
\begin{equation}
\label{eq:scalarsgeneral}
z^i=\frac{\mathcal{X}^{i}}{\mathcal{X}^0} \ \, .
\end{equation}
\noindent
Therefore, this coordinate system is only valid away from the locus $\mathcal{X}^{0}=0$.

We are interested in studying spherically symmetric, static, black hole solutions of the theory defined by Eq. (\ref{eq:IIaprepotential}). In order to do so we are going to use the so-called H-FGK, developed in \cite{Galli:2011fq,Meessen:2011aa,Meessen:2012su}, based on the use of a new set of variables $H^M,~~ M=(\Lambda,\Lambda)$, which transform linearly under duality and reduce to harmonic functions on the transverse space $\mathbb{R}^{3}$ in the supersymmetric case. 



\section{A non-perturbative class of black holes}
\label{sec:blackholes}


The most general static, spherically symmetric space-time metric solution of an ungauged Supergravity is given by\footnote{The conformastatic coordinates $\left(t,\tau,\theta, \phi\right)$ cover the outer region of the event horizon when $\tau\in\left(-\infty, 0\right)$ and the inner region, between the Cauchy horizon and the physical singularity when $\tau\in\left(\tau_S, \infty\right)$, where $\tau_S\in\mathbb{R}^{+}$ is a model dependent number. The event horizon is located at $\tau \rightarrow -\infty$ and the Cauchy horizon at $\tau\rightarrow \infty$.} \cite{Ferrara:1997tw,Meessen:2011aa}
\begin{equation}
\label{eq:generalbhmetric}
\begin{array}{rcl}
ds^{2}
& = &
e^{2U(\tau)} dt^{2} - e^{-2U(\tau)} \gamma_{\underline{m}\underline{n}}
dx^{\underline{m}}dx^{\underline{n}}\, ,  \\
& & \\
\gamma_{\underline{m}\underline{n}}
dx^{\underline{m}}dx^{\underline{n}}
& = &
{\displaystyle\frac{r_{0}^{4}}{\sinh^{4} r_{0}\tau}}d\tau^{2}
+
{\displaystyle\frac{r_{0} ^{2}}{\sinh^{2}r_{0}\tau}}d\Omega^{2}_{(2)}\, .\\
\end{array}
\end{equation}

\noindent
Using Eq. (\ref{eq:generalbhmetric}), and assuming spherical symmetry for all the fields in the theory, the equations of motion of the bosonic sector of $\mathcal{N}=2$, $d=4$ Supergravity can be written as a set of ordinary differential equations (in the variable $\tau$) for the scalars $z^i(\tau)$, $i=1,...,n_v$ and the metric warp factor $U(\tau)$ \cite{Ferrara:1997tw}. The vector fields do not appear in these equations since they can be explicitly integrated in terms of the corresponding electric and magnetic charges\footnote{The form of the vector fields can be recovered following the dimensional-reduction procedure. Let $\Psi = (\psi^{\Lambda},\chi_{\Lambda})^{T}$ be a symplectic vector whose components are the time components of the electric $A^{\Lambda}$ and magnetic $A_{\Lambda}$ vector fields. Then, $\Psi$ is given by

\begin{equation}
\Psi = \int\frac{1}{2} e^{2U} \mathcal{M}^{MN}\mathcal{Q}_{N}d\tau\, ,
\end{equation}
\noindent
where $\mathcal{M}^{MN}$ is a symplectic matrix constructed from the couplings of the scalars and the vector fields. For more details see \cite{Simone}.}.

The one dimensional effective equations of motion can be recast in a duality covariant way by performing a particular change of variables to a new set of $(2n_v+2)$ variables $H^M(\tau)$ which transform linearly under the U-duality group of the theory, and become harmonic functions in $\mathbb{R}^3$ in the supersymmetric case. This is the essence of the H-FGK formalism \cite{Galli:2011fq,Meessen:2011aa,Meessen:2012su,Galli:2012ji}, whose equations of motion read\\ ( $\dot{}\equiv \frac{d}{d\tau}$)
\begin{align}
\label{eq:Eqsofmotion}
\mathcal{E}_{P}=\tfrac{1}{2}\partial_{P}\partial_{M}\partial_{N}\log \mathsf{W}\,
\left[\dot{H}^{M}\dot{H}^{N} -\tfrac{1}{2}\mathcal{Q}^{M}\mathcal{Q}^{N}
\right]
+\partial_{P}\partial_{M}\log \mathsf{W}\, \ddot{H}^{M}
-\frac{d}{d\tau}\left(\frac{\partial \Lambda}{\partial \dot{H}^{P}}\right)
+\frac{\partial \Lambda}{\partial H^{P}}=0\, ,
\end{align}

\noindent
together with the \emph{Hamiltonian constraint}
\begin{align}
\label{eq:Hamiltonianconstraint}
\mathcal{H}\equiv
-\tfrac{1}{2}\partial_{M}\partial_{N}\log\mathsf{W}
\left(\dot{H}^{M}\dot{H}^{N}-\tfrac{1}{2}\mathcal{Q}^{M}\mathcal{Q}^{N} \right)
+\left(\frac{\dot{H}^{M}H_{M}}{ \mathsf{W}}\right)^{2}
-\left(\frac{\mathcal{Q}^{M}H_{M}}{ \mathsf{W}}\right)^{2}
-r_{0}^{2}=0\, ,
\end{align}

\noindent
where
\begin{equation}
\Lambda \equiv \left(\frac{\dot{H}^{M}H_{M}}{ \mathsf{W}}\right)^{2}
+\left(\frac{\mathcal{Q}^{M}H_{M}}{ \mathsf{W}}\right)^{2}\, ,
\end{equation}

\noindent
and
\begin{equation}
\label{eq:W(H)}
\mathsf{W}(H) \equiv \tilde{H}_{M}(H)H^{M} = e^{-2U}, ~~~\tilde{H}^M+i\mathcal{I}^M =\mathcal{V}^M/X\, .
\end{equation}

\noindent
$\mathcal{V}^M$ is the covariantly holomorphic symplectic section of $\mathcal{N}=2$ Supergravity, and $X$ is a complex variable with the same K\"ahler weight as $\mathcal{V}^M$. $\tilde{H}^M(\mathcal{I})\equiv \tilde{H}^M(H)$ stands for the real part $\left(\tilde{H}^M\right)$ of $\mathcal{V}^M$ written as a function of the imaginary part $\mathcal{I}^M\equiv H^M$, something that can always be done by solving the so-called \emph{stabilization equations}, which we will explicitly write down in a moment. $\mathsf{W}(H)$ is usually known in the literature as the \emph{Hesse potential}.

The effective theory is now expressed in terms of $2\left(n_{v}+1\right)$ variables $H^M$ and depends on $2\left(n_{v}+1\right)+1$ parameters: $2\left(n_{v}+1\right)$ charges $\mathcal{Q}^M=\left(p^{\Lambda},\, \, q_{\Lambda}\right)^T$ and the non-extremality parameter $r_0$, from which it is possible to reconstruct the solution in terms of the original fields of the theory (that is it, the space-time metric, scalars and vector fields).

In order to tackle the construction of new black hole solutions of (\ref{eq:IIaprepotential}), we are going to consider a particular consistent truncation given by

\begin{equation}
\label{eq:truncation}
H^0 = H_0 = H_i = 0,~~p^{0} = q_{0} = q_{i} = 0 \, .
\end{equation}

\noindent
Under this assumption, the stabilization equations, which can be directly read off from (\ref{eq:W(H)}) take the form

\begin{equation}
\label{eq:stab}
\left( \begin{array}{c} iH^i \\\tilde{H}_i  \end{array} \right)=\frac{e^{\mathcal{K}/2}}{X}\left( \begin{array}{c} \mathcal{X}^i \\ \frac{\partial F(\mathcal{X})}{\partial\mathcal{X}^i}  \end{array} \right)
 \, ,
\left( \begin{array}{c} \tilde{H}^0\\0  \end{array}\right)=\frac{e^{\mathcal{K}/2}}{X}\left( \begin{array}{c} \mathcal{X}^0 \\ \frac{\partial F(\mathcal{X})}{\partial\mathcal{X}^0}  \end{array} \right),
\end{equation}
\noindent
where we have used that the covariantly holomorphic symplectic section $\mathcal{V}^M$ can be written as follows \cite{Andrianopoli:1996cm}
\begin{equation}
\label{v}
\mathcal{V}^M=e^{\mathcal{K}/2}\left(\mathcal{X}^{\Lambda},\, \, \partial_{\Lambda}F(\mathcal{X})\right)\, .
\end{equation}

The physical fields can be obtained from the $H^i$ as
\begin{equation}
\label{eq:phy}
e^{-2U}=\tilde{H}_iH^i \, ,~~z^i=i\frac{H^i}{\tilde{H}^0}\, ,
\end{equation}

\noindent
as soon as $\tilde{H}^0$ and $\tilde{H}^i$ are determined. In order to obtain $\tilde{H}^0$ as a function of $H^i$, we need to solve the highly involved equation

\begin{equation}
\label{eqR0}
\frac{\partial{F(H)}}{\partial \tilde{H}^0}=0\, ,
\end{equation}

\noindent
where $F(H)$ stands for the prepotential expressed in terms of the $H^i$
\begin{align}
\label{eq:IIaprepotentialH}
F(H) =  \frac{i}{3!}\kappa^{0}_{ijk} \frac{H^i H^j H^k}{\tilde{H}^0} +\frac{ic(\tilde{H}^0)^2}{2}+\frac{i(\tilde{H}^0)^2}{(2\pi)^3}\sum_{\{d_{i}\}} n_{\{d_{i}\}} Li_{3}\left(e^{-2\pi  d_{i} \frac{H^{i}}{\tilde{H}^0}}\right)  \, .
\end{align}

\noindent
Once this is done, it is not difficult to express $\tilde{H}^i$ in terms of $H^i$. Indeed, from (\ref{eq:stab}) we simply have

\begin{equation}
\label{eq:ri}
\tilde{H}_i=-i\frac{\partial{F(H)}}{\partial H^i}\, .
\end{equation}

\noindent
If we expand (\ref{eqR0}), we find
\begin{align}
\label{d}
  -\frac{1}{3!}\kappa^{0}_{ijk} \frac{H^i H^j H^k}{({\tilde{H}^0})^3} +c+\frac{1}{4\pi^3}\sum_{\{d_{i}\}} n_{\{d_{i}\}} \left[Li_{3}\left(e^{-2\pi  d_{i} \frac{H^{i}}{\tilde{H}^0}}\right) \right. \left. +Li_{2}\left(e^{-2\pi  d_{i} \frac{H^{i}}{\tilde{H}^0}}\right)\left[\frac{\pi d_iH^i}{\tilde{H}^0} \right] \right]=0\, .
\end{align}

\noindent
Solving (\ref{d}) for $\tilde{H}^0$ in full generality seems to be an extremely difficult task. However, if we go to the large volume compactification limit ($\Im{\rm m} z^i >> 1$), we can make use of the following property of the polylogarithmic functions
\begin{equation}
\label{eq:lilimit}
\lim_{|w|\rightarrow 0}Li_s(w)=w\, , \forall s\in \mathbb{N}\, ,
\end{equation}

\noindent
since, in our case, $w=e^{-2\pi d_{i}\Im m z^{i}}\, , \forall \left\{d_i\right\}\in\left(\mathbb{Z}^{+}\right)^{h^{1,1}}$. Eq. (\ref{eq:lilimit}) enables us to rewrite (\ref{d}) as

\begin{align}
\label{di}
  -\frac{1}{3!}\kappa^{0}_{ijk} \frac{H^i H^j H^k}{({\tilde{H}^0})^3} +c+\frac{1}{4\pi^3}\sum_{\{d_{i}\}} n_{\{d_{i}\}} \left[e^{-2\pi  d_{i} \frac{H^{i}}{\tilde{H}^0}}\right.  \left.+e^{-2\pi  d_{i} \frac{H^{i}}{\tilde{H}^0}}\left[\frac{\pi d_iH^i}{\tilde{H}^0} \right] \right]= 0\, , ~~\Im{\rm m}z^{i} >> 1\, .
\end{align}

\noindent
The dominant contribution in this regime, aside from the cubic one, is given by $c$. In \cite{Bueno:2012jc} \cite{Galli:2012pt}, the first non-extremal black hole solutions (with constant and non-constant scalars) of (\ref{eq:IIaprepotential}) were obtained ignoring the non-perturbative corrections. In particular, the solutions of \cite{Bueno:2012jc} turned out to be purely \textit{quantum}\footnote{It is worth pointing out again that the term \textit{quantum} does not refer to space-time but to world-sheet properties in this context \cite{Mohaupt:2000mj}. In this respect, although such denomination is widely spread in the literature, the adjective \textit{stringy} might result more acqurate.}, in the sense that not only the classical limit $c\rightarrow 0$ was ill-defined, but also the truncated theory became inconsistent and therefore no classical limit could be assigned to such solutions. An interesting question to ask now is whether the non-perturbative contributions could actually be able to cure or at least improve this behaviour. On the other hand, it is also interesting \textit{per se} to explore the existence of black hole solutions when the subleading contribution to the prepotential is not given by $c$, but has a non-perturbative origin. In order to tackle these two questions, let us restrict ourselves to C.Y. three-folds with vanishing Euler characteristic ($c=0$), the so-called self-mirror C.Y. three-folds. Under this assumption, and considering only the subleading contribution in (\ref{d}), which is now given by the fourth term in (\ref{di}), such equation becomes\footnote{$e^{2\pi i d_{i}z^i}<< \pi d_{i}\Im{\rm m}z^i e^{2\pi i d_{i}z^i}$ for $\Im{\rm m}z^{i} >> 1$.}

\begin{equation}
\label{dii}
  -\frac{1}{3!}\kappa^{0}_{ijk} \frac{H^i H^j H^k}{({\tilde{H}^0})^3}+\frac{1}{4\pi^3}\sum_{\{d_{i}\}} n_{\{d_{i}\}} e^{-2\pi  d_{i} \frac{H^{i}}{\tilde{H}^0}}\left[\frac{\pi d_iH^i}{\tilde{H}^0} \right] = 0\, .
\end{equation}

\noindent
The sum over $\{d_i\}$ in (\ref{dii}) will be dominated in each case by a certain term corresponding to a particular vector $\left\{\hat{d}_i\right\}$ 
(and, as a consequence, to a particular $n_{\hat{d}_{i}}\equiv \hat{n}$), which, since we are assuming $\Im{\rm m}z^{i}>>1$, is the only one that we need to consider. That is, $\left\{\hat{d}_{i}\right\}$ corresponds to the set of $d_{i}$ that labels the most relevant term in the infinite sum present in (\ref{dii}). Hence, this equation becomes

\begin{equation}
\label{diii}
  -\frac{1}{3!}\kappa^{0}_{ijk} \frac{H^i H^j H^k}{({\tilde{H}^0})^3}+\frac{\hat{n}}{4\pi^3}  e^{-2\pi  \hat{d}_{i} \frac{H^{i}}{\tilde{H}^0}}\left[\frac{\pi \hat{d}_iH^i}{\tilde{H}^0} \right] = 0\, ,~~ \Im{\rm m}z^{i} >> 1\, .
\end{equation}

\noindent
This is solved by\footnote{Henceforth we will be using $\mathsf{W}$ for the Hessian potential, and $W$ for the Lambert function. We hope this is not a source of confusion.}
\begin{equation}
\label{roW}
\tilde{H}^0=\frac{\pi \hat{d}_lH^l}{ W_a\left(s_a \sqrt{\frac{3\hat{n}(\hat{d}_nH^n)^3}{2\kappa^0_{ijk}H^i H^j H^k}}\right)} \, ,
\end{equation}

\noindent
where $W_a(x), ~(a=0,-1)$ stands for (any of the two real branches of) the \textit{Lambert $W$ function}\footnote{See the Appendix \ref{sec:lambert} for more details.} (also known as \textit{product logarithm}), and $s_a=\pm1$. Using now Eqs. (\ref{roW}) and (\ref{eq:ri}) we can obtain $\tilde{H}^i$. The result is


\begin{equation}
\tilde{H}_i=\frac{1}{2}\kappa^{0}_{ijk} \frac{H^j H^k}{\pi \hat{d}_lH^l}W_a\left(s_a \sqrt{\frac{3\hat{n}(\hat{d}_mH^m)^3}{2\kappa^0_{pqr}H^p H^q H^r}}\right)\, .
\end{equation}

\noindent
The physical fields can now be written as a function of the $H^i$ as

\begin{equation}\label{hessian}
e^{-2U}=\mathsf{W}(H)=\frac{\kappa^{0}_{ijk} H^i H^j H^k}{2\pi \hat{d}_m H^m} W_a\left(s_a\sqrt{\frac{3\hat{n}(\hat{d}_lH^l)^3}{2\kappa^0_{pqr}H^p H^q H^r}}\right)\, ,
\end{equation}
\begin{equation}\label{scalar}
z^i=i\frac{H^i}{\pi \hat{d}_mH^m}W_a\left(s_a \sqrt{\frac{3\hat{n}(\hat{d}_lH^l)^3}{2\kappa^0_{pqr}H^p H^q H^r}}\right)\, .
\end{equation}

\noindent
In order to have a regular solution, we need to have a positive definite metric warp factor $e^{-2U}$. Since, as explained in Appendix \ref{sec:lambert}, ${\rm sign} \left[W_a(x)\right]={\rm sign}\left[x\right],~a=0,-1,~x\in D^a_{\mathbb{R}}$,  we have to require that
\begin{equation}
s_0\equiv sign\left[\kappa^{0}_{ijk} \frac{H^i H^j H^k}{ \hat{d}_m H^m} \right]\, ,
\end{equation}

\begin{equation}
s_{-1}\equiv -1\, .
\end{equation}

\noindent
On the other hand, since $W_{0}(x)=0$ for $x=0$ and $W_{-1}(x)$ is a real function only when $x\in\left[-\frac{1}{e},0\right)$, we have to impose that the argument $x$ of $W_{a}$ lies entirely either in $\left[-\frac{1}{e},0\right)$ or in $\left(0,\infty\right)$ for all $\tau\in\left(-\infty,0\right)$, since $e^{2U}$ cannot be zero in a regular black hole solution for any $\tau\in\left(-\infty,0\right)$. This condition must be imposed in a case by case basis, since it depends on the specific form of the symplectic vector $H^M=H^M (\tau)$ as a function of $\tau$. Notice that if $x\in \left[-\frac{1}{e},0\right)~~\forall~~\tau\in\left(-\infty,0\right)$ we can in principle\footnote{As we will see in section \ref{sec:susysolution}, the possibility $s_0=s_{-1}=-1$ will not be consistent with the large volume approximation we are considering.} choose either $W_{0}$ or $W_{-1}$ to build the solution, whereas if $x\in \left(0,+\infty \right)~~\forall~~\tau\in\left(-\infty,0\right)$, only $W_{0}$ is available.

Needless to say, in order to construct actual solutions, we have to solve the H-FGK equations of motion (\ref{eq:Eqsofmotion}) (plus hamiltonian constraint (\ref{eq:Hamiltonianconstraint})) using the Hessian potential given by (\ref{hessian}). Fortunately, such equations admit a \emph{model-independent} solution which is obtained choosing the $H^{i}$ to be harmonic functions in the flat transverse space, with one of the poles given in terms of the corresponding charge
\begin{equation}
\label{eq:universalsusy}
H^{i} = a^{i}-\frac{p^{i}}{\sqrt{2}}\tau,~~~~r_0=0\, .
\end{equation}

\noindent
In fact, it is a virtue of the H-FGK formalism to make explicit how every $\mathcal{N}=2$, $d=4$ Supergravity theory admits a solution of the form
\begin{equation}
\label{eq:universalsusyi}
H^{M} = a^{M}-\frac{\mathcal{Q}^{M}}{\sqrt{2}}\tau,~~~~r_0=0\, ,~~~~a^M\mathcal{Q}_M=0\, , 
\end{equation} 
\noindent
where the last equation encodes the absence of Taub-NUT charge. It can be easily verified that Eq. (\ref{eq:universalsusyi}) does indeed satisfy Eqs. (\ref{eq:Eqsofmotion}) and (\ref{eq:Hamiltonianconstraint}) independently of the model.
This corresponds to a supersymmetric black hole solution \cite{Tod:1983pm,Behrndt:1997ny, Meessen:2006tu}.


\section{The general supersymmetric solution}
\label{sec:susysolution}


As we have said, plugging (\ref{eq:universalsusy}) into (\ref{scalar}) and (\ref{hessian}) provides us with a supersymmetric solution without solving any further equation. The entropy of such a solution reads
\begin{equation}
S=\frac{1}{2}\kappa^{0}_{ijk} \frac{p^i p^j p^k}{ \hat{d}_m p^m} W_a\left(s_a \beta\right)\, ,
\end{equation}
\begin{equation}\notag
\beta=\sqrt{\frac{3\hat{n}(\hat{d}_l p^l)^3}{2\kappa^0_{pqr}p^p p^q p^r}}\, ,
\end{equation}

\noindent
and the mass is given by
\begin{align}
\label{mass}
M =\dot{U}(0)=\frac{1}{2\sqrt{2}}\left[\frac{3\kappa^0_{ijk}p^i a^j a^k}{\kappa^0_{pqr}a^p a^q a^r}\left[1-\frac{1}{1+W_a(s_a\alpha)} \right] \right.\left.-\frac{d_l p^l}{d_n a^n}\left[1-\frac{3}{2\left(1+W_a(s_a\alpha)\right)} \right] \right] \, ,
\end{align}
\begin{equation}
\alpha= \sqrt{\frac{3\hat{n}(\hat{d}_l a^l)^3}{2\kappa^0_{pqr}a^p a^q a^r}} \, .
\end{equation}

\noindent
In the approximation under consideration, we are neglecting terms $\sim e^{-2\pi d_i \Im m z^i}$ with respect to those going as $\sim \pi d_i \Im m z^i e^{-2\pi d_i \Im m z^i}$. Taking into account (\ref{scalar}), this assumption is translated into the condition

\begin{equation}
\label{condon}
W_a(x) e^{-2 W_a(x)}>>e^{-2 W_a(x)}\, .
\end{equation}

\noindent
It is clear that this condition is satisfied for $a=0$ if $x\in [\alpha,\beta]$ for positive and suficiently large values of $\alpha$ and $\beta$. Constructing a solution such that the values of $p^i$ and $a^i$ correspond to large enough $\alpha$ and $\beta$ may or may not be possible depending on the compactification data. For example, if $\hat{d_i}=(1,0,...,0)$ and $\kappa^0_{iii}=0\, \forall\, i=1,...,n_v$, it is clear that taking $a^1>>1$ and $p^1>>1$ satisfies the corresponding condition.

 It is also clear, however, that (\ref{condon}) is not satisfied at all for $x\in [-\frac{1}{e},0)$, which is the range for which both branches of the Lambert function are available. 

If we assume $x \in \left[\alpha,\beta \right]$ for suficiently large $\alpha,\beta\in\mathbb{R}^+$, $a=0$ and $W_0$ is the only real branch of the Lambert function. In that case, $s=s_0=1$, and we have
\begin{equation}\label{hessian0}
e^{-2U}=\frac{\kappa^{0}_{ijk} H^i H^j H^k}{2\pi \hat{d}_m H^m} W_0\left(\sqrt{\frac{3\hat{n}(\hat{d}_lH^l)^3}{2\kappa^0_{pqr}H^p H^q H^r}}\right)\, ,
\end{equation}
\begin{equation}\label{scalar0}
z^i=i\frac{H^i}{\pi \hat{d}_mH^m}W_0\left( \sqrt{\frac{3\hat{n}(\hat{d}_lH^l)^3}{2\kappa^0_{pqr}H^p H^q H^r}}\right)\, .
\end{equation}

\noindent
In the conformastatic coordinates we are working with, the metric warp factor $e^{-2U}$ is expected to diverge at the event horizon ($\tau\rightarrow -\infty$) as $\tau^2$. In addition, we have to require $e^{-2U}>0$ $\forall \tau \in (-\infty,0]$, and impose asymptotic flatness $e^{-2U(\tau=0)}=1$. The last two conditions read

\begin{equation}\label{posimetri}
\frac{\kappa^0_{ijk} H^i H^j H^k}{2\pi \hat{d}_n H^n}>0 ~~\forall \tau\in (-\infty,0] \, ,
\end{equation}
\begin{equation}\label{asyfla}
 \frac{\kappa^0_{ijk}a^i a^j a^k}{2\pi \hat{d}_m a^m}W_0\left(\alpha \right)=1\, ,
\end{equation}

\noindent
whereas the first one turns out to hold, since
\begin{equation}
e^{-2U}\overset{\tau \rightarrow -\infty}{\longrightarrow} \frac{\kappa^0_{ijk}p^i p^j p^k}{8\pi \hat{d}_m p^m}W_0\left(\beta\right)\tau^2  \,.
\end{equation}

\noindent
(\ref{posimetri}) and (\ref{asyfla}) can in general be safely imposed in any particular model we consider. Finally, the condition for a well-defined and positive mass $M>0$ can be read off from (\ref{mass}).

\section{Multivalued functions and the No-Hair conjecture}

As we explained in the previous section, our approximation is not consistent with a solution such that $x\in [-\frac{1}{e},0)$. This forbids the domain in which $W(x)$ is a multivalued function (both $W_0$ and $W_{-1}$ are real there). However, it seems legitimate to ask what the consequences of having two different branches would have been, had this constraint not been present.
In principle, we could have tried to assign the asymptotic ($\tau \rightarrow 0$) and near horizon ($\tau \rightarrow -\infty$) limits to any particular pair of values of the arguments of $W_0$ and $W_{-1}$ ($x_0$ and $x_{-1}$ respectively) through a suitable election of the parameters available in the solution. In particular, had we chosen $x_{0}|_{\tau=0}=x_{-1}|_{\tau=0}=-1/e$ and $x_{0}|_{\tau \rightarrow -\infty}=x_{-1}|_{\tau \rightarrow -\infty}=\beta$, $\beta \in (-1/e,0)$, both solutions would have had exactly the same asymptotic behavior (and therefore the scalars of both solutions would have coincided at spatial infinity), and we would have been dealing with two completely different regular solutions with the same mass\footnote{Although $W^{\prime}_{0,-1}(x)$ are divergent at $x=-1/e$ (as explained in the Appendix \ref{sec:lambert}), and the definition of $M$ would involve derivatives of the Lambert function at that point, it would not be difficult to cure this behaviour and get a positive (and finite) mass by imposing $\dot{x}(\tau)\overset{\tau \rightarrow 0 }{\longrightarrow 0}$ faster than $|W^{\prime}_{0,-1}(x)|\overset{x \rightarrow -1/e}{\longrightarrow \infty}$.}, charges and asymptotic values of the scalar fields, in flagrant contradiction\footnote{Up to possible stability issues, which should be carefully studied.} with the corresponding black hole uniqueness conjecture (and, as a consequence, with the No-Hair conjecture). At this point, and provided that our approximation is not consistent with such presumable two-branched solution, the feasibility of this reasoning in a different context can only be catalogued as \textit{speculative} at the very least. However, a violation of the No-Hair conjecture in four dimensions would have far-reaching consequences independently of whether the solution is embedded in String Theory or not. In this regard, the very possibility that the stabilization equations may admit (for certain more or less complicated prepotentials) solutions depending on multivalued functions seems to open up a window for possible violations of the No-Hair conjecture in the context of $\mathcal{N}=2$ $d=4$ Supergravity. The question (whose answer is widely assumed to be "no") is now: is it possible to find a four-dimensional (Super)gravity theory with a physically-admisible matter content admitting more than one stable black hole solution with the same mass, electric, magnetic and scalar charges? If not, why? These questions will be addressed in a forthcoming publication \cite{POS}.

\section*{Acknowledgments}


We wish to thank T. Ort\'in and D. Regalado for very useful discussions and A. Kirillov for useful comments. We are also grateful to our team of French translators: P. Fleury, C. G. Alc\'antara and S. C. Moure. Finally, we are thankful to Wisin \& Yandel for very enjoyable moments during the work sessions. This work has been supported in part by the Spanish Ministry of Science and Education grant FPA2012-35043-C02-01, the Comunidad de Madrid grant HEPHACOS S2009ESP-1473, and the Spanish Consolider-Ingenio 2010 program CPAN CSD2007-00042. The work of PB and CSS has been supported by the JAE-predoc grants JAEPre 2011 00452 and JAEPre 2010 00613 respectively.

\appendix


\section{The polylogarithm}
\label{sec:polylog}


The \textit{polylogarithmic function} or \textit{polylogarithm} $Li_w(z)$ (see e.g. \cite{Lewin} for an exhaustive study) is a special function defined through the power series

\begin{equation}
Li_w(z)=\sum_{j=1}^{\infty} \frac{z^j}{j^w}\, ,~~z,w\in \mathbb{C}\, .
\end{equation}

\noindent
This definition is valid for arbitrary complex numbers $w$ and $z$ for $|z|<1$, but can be extended to $z's$ with $|z|\geq 1$ by analytic continuation. From its definition, it is easy to find the recurrence relation

\begin{equation}
Li_{w-1}(z)=z\frac{\partial Li_w(z)}{\partial z}\, .
\end{equation}
\noindent
The case $w=1$ corresponds to

\begin{equation}
Li_1(z)=-\log(1-z) \, ,
\end{equation}

\noindent
and from this it is easy to see that for $w=-n\in \mathbb{Z}^- \cup \left\{0 \right\}$, the polylogarithm is an elementary function given by

\begin{equation}
Li_0(z)=\frac{z}{1-z}\, ,~~ Li_{-n}(z)=\left(z\frac{\partial}{\partial z} \right)^n \frac{z}{1-z}\, .
\end{equation}

\noindent
The special cases $w=2,3$ are called \textit{dilogarithm} and \textit{trilogarithm} respectively, and their integral representations can be obtained from $Li_1(z)$ making use of

\begin{equation}
Li_{w}(z)=\int^{z}_0 \frac{Li_{w-1}(s)}{s}ds\, .
\end{equation}



\section{The Lambert W function}
\label{sec:lambert}


The \textit{Lambert W function} $W(z)$ (also known as \textit{product logarithm}) is named after Johann Heinrich Lambert (1728-1777), who was the first to introduce it in 1758 \cite{Lambert}. During its more than two hundred years of history, it has found numerous applications in different areas of physics (mainly during the 20th century) such as electrostatics, thermodynamics (e.g. \cite{Valluri}), statistical physics (e.g. \cite{JM}), QCD (e.g. \cite{Gardi:1998qr,Magradze:1998ng,Nesterenko:2003xb,Cvetic:2011vy,Sonoda:2013kia}), cosmology (e.g. \cite{Ashoorioon:2004vm}), quantum mechanics (e.g. \cite{Mann}) and general relativity (e.g. \cite{Mann:1996cb}). 

$W(z)$ is defined implicitly through the equation 

\begin{equation}
\label{Wf}
z=W(z)e^{W(z)}\, ,~~ \forall z\in \mathbb{C}\,.
\end{equation}

\noindent
Since $f(z)=ze^{z}$ is not an injective mapping, $W(z)$ is not uniquely defined, and $W(z)$ generically stands for the whole set of branches solving (\ref{Wf}). For $W:\mathbb{R}\rightarrow \mathbb{R}$, $W(x)$ has two branches $W_0(x)$ and $W_{-1}(x)$ defined in the intervals $x\in [-1/e,+\infty)$ and $x\in [-1/e,0)$ respectively (See Figure 1). Both functions coincide in the branching point $x=-1/e$, where $W_0(-1/e)=W_{-1}(-1/e)=-1$. As a consequence, the defining equation $x=W(x)e^{W(x)}$ admits two different solutions in the interval $x\in [-1/e,0)$.\\
\begin{figure}[h]
 \label{fig:u}
  \centering
    \includegraphics[scale=0.55]{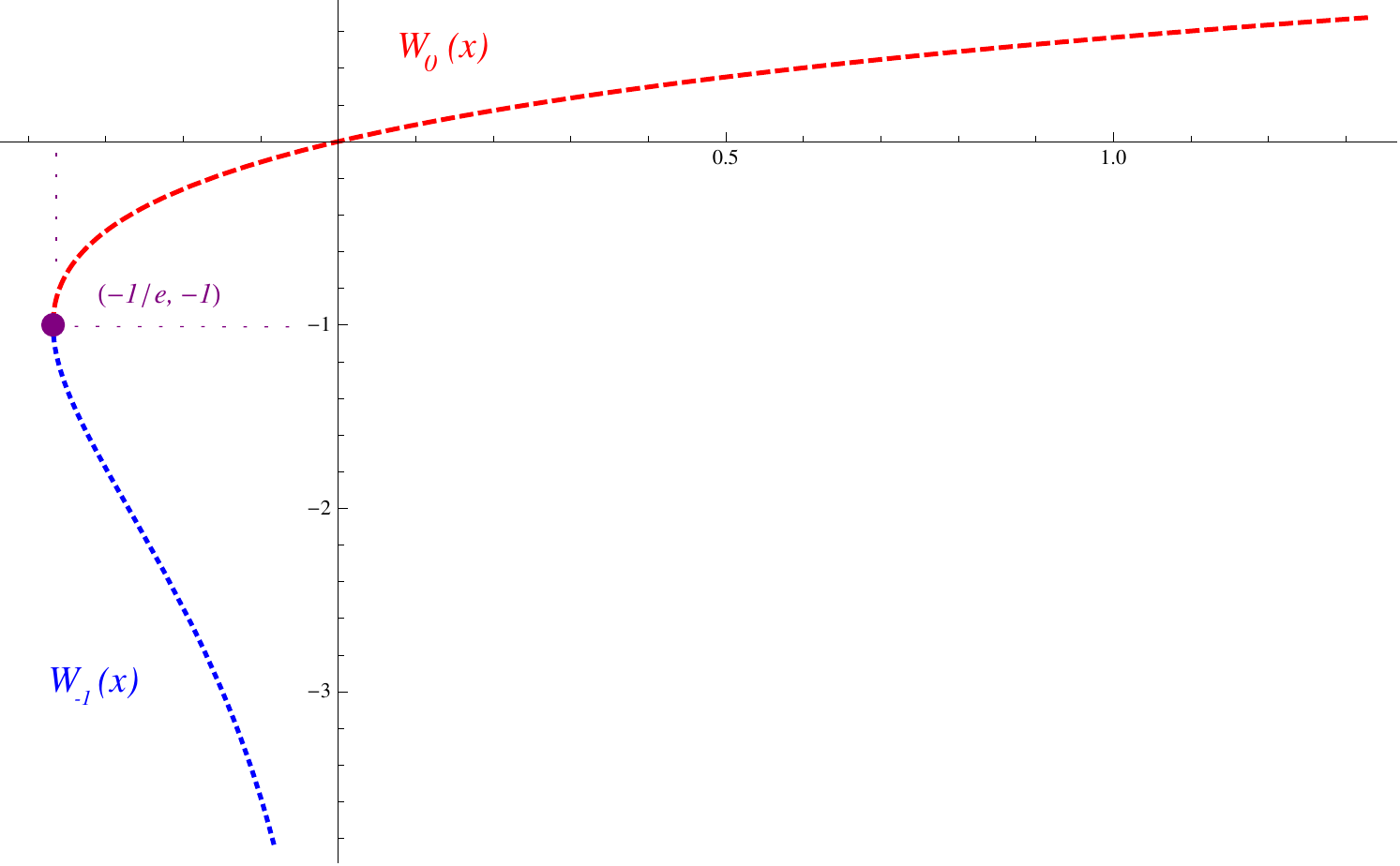}
 \caption{\small{The two real branches of $W(x)$.}}
\end{figure}

\noindent
The derivative of $W(z)$ reads

\begin{equation}
\frac{dW(z)}{dz}=\frac{W(z)}{z(1+W(z))},~~\forall z \notin \left\{0,-1/e \right\}; ~~ \frac{dW(z)}{dz}\bigg|_{z=0}=1 \, ,
\end{equation}

\noindent
and is not defined for $z=-1/e$ (the function is not differentiable there). At that point we have

\begin{equation}
\lim_{x\rightarrow -1/e}\frac{dW_0(x)}{dx}=\infty,~~\lim_{x\rightarrow -1/e}\frac{dW_{-1}(x)}{dx}=-\infty \, .
\end{equation}

\renewcommand{\leftmark}{\MakeUppercase{Bibliography}}
\phantomsection
\addcontentsline{toc}{chapter}{References}
\bibliographystyle{JHEP}
\bibliography{References}
\label{biblio}

\line(1,0){100}

\noindent
Instituto de F\'isica Te\'orica UAM/CSIC,\\ C/ Nicol\'as Cabrera, 13-15, C.U. Cantoblanco, 28049 Madrid, Spain\\\\
\textit{IFT Number}: IFT-UAM/CSIC-13-054\\\\
\textit{E-mail}:\\
 $\textcolor{Red}{\varheart}$\texttt{pab.bueno[at]estudiante.uam.es\\ $\textcolor{Purple}{\varheart}$carlos.shabazi[at]uam.es}\\
\noindent
\input Starburst.fd
\newcommand*\initfamily{\usefont{U}{Starburst}{xl}{n}}
\end{document}